\documentclass[a4paper]{jpconf}

\usepackage{amsmath,amssymb}
\usepackage{epsfig}
\usepackage{graphicx,color}
\usepackage[sort&compress]{natbib}

\bibpunct{[}{]}{,}{n}{}{,}

\allowdisplaybreaks

\newcommand{\eps}{{\ensuremath{\epsilon}}}

\newcommand{\beq}{\begin{equation}}
\newcommand{\eeq}{\end{equation}}

\begin{document}

%==============================================================================
\title{New signals in dark matter detectors}
\author{Joachim Kopp}
\address{Fermilab, PO Box 500, Batavia, IL 60510,USA and \\
         Max--Planck--Institut f\"{u}r Kernphysik,
         Postfach 10 39 80, 69029 Heidelberg, Germany}
%\author{J Kopp$^{1,2}$, P Machado$^{3,4}$, R Harnik$^1$}
%\address{$^1$ Fermilab, PO Box 500, Batavia, IL 60510,USA}
%\address{$^2$ Max--Planck--Institut f\"{u}r Kernphysik,
%              Postfach 10 39 80, 69029 Heidelberg, Germany}
%\address{$^3$ Instituto de F\'{i}sica, Universidade de S\~{a}o Paulo,
%              C.P.~66.318, 05315-970 S\~{a}o Paulo, Brazil}
%\address{$^4$ Institut de Physique Th\'{e}orique, CEA Saclay,
%              91191 Gif-sur-Yvette, France}

\ead{jkopp@mpi-hd.mpg.de}
%==============================================================================

\begin{abstract}
We investigate the scattering of solar neutrinos on electrons and nuclei in
dark matter direct detection experiments. The rates of these processes are
small in the Standard Model, but can be enhanced by several orders of magnitude
if the neutrino sector is slightly non-minimal. This makes even the current
generation of dark matter detectors very sensitive to non-standard neutrino
physics. Examples discussed here are neutrino magnetic moments and toy models
with a simple hidden sector containing a sterile neutrino and a light new gauge
boson (``dark photon'').  We discuss the expected event spectra and temporal
modulation effects, as well as constraints from a variety of astrophysical,
cosmological, and laboratory experiments.
\end{abstract}

%------------------------------------------------------------------------------
\section{Neutrino interactions in dark matter detectors}
%------------------------------------------------------------------------------

It is well known that solar and atmospheric neutrinos constitute an irreducible
background to future dark matter searches (see for
instance~\cite{Gutlein:2010tq}).  In particular, neutrinos can scatter on
atomic nuclei or electrons, thus mimicking typical dark matter signatures. In
this talk, based mostly on ref.~\cite{Harnik:2012ni}, we assume that the
neutrino sector is slightly non-minimal and show that, in this case, neutrino
signals in dark matter detectors can be up to several orders of magnitude
stronger than in the Standard Model, enough to make them highly relevant even
in the current generation of experiments. (This is also discussed in
refs.~\cite{Pospelov:2011ha, Pospelov:2012gm}.)

At the basis of our discussion lies the observation that interactions mediated
by particles much lighter than the typical momentum transfers in the process
become stronger at low energy. Hence, if such interactions exist in the neutrino
sector, they may be very significant in dark matter detectors with their low
energy thresholds of 10~keV, while dedicated neutrino detectors, whose energy
thresholds are at least a few hundred keV would be insensitive to them.

We consider here two examples for light new physics in the neutrino sector:
A neutrino magnetic moment (where the light mediator is the photon) and
scenarios with a gauged sterile neutrino sector (where the light mediator is a
new $U(1)'$ gauge boson).

\subsection{Neutrino magnetic moments}
%-------------------------------------

Neutrino magnetic moments are described by a Lagrangian operator of the form
\begin{align}
  \mathcal{L}_{\mu_\nu} \supset \mu_\nu \, \bar\nu
    \sigma^{\alpha\beta} \partial_\beta A_\alpha \nu \,,
  \label{eq:L-mm}
\end{align}
where $A_\alpha$ and $\nu$ denote the photon and neutrino fields, respectively,
$\mu_\nu$ is the neutrino magnetic moment, and as usual, $\sigma^{\alpha\beta} =
\frac{i}{2} [\gamma^\alpha, \gamma^\beta]$. The Standard Model prediction for
$\mu_\nu$ is negligibly small~\cite{Beringer:2012zz}, but in extensions of the
Standard Model, it can come close to the current experimental 90\%~CL upper
limit $\mu_\nu < 0.32 \times 10^{-10} \mu_B$ (where $\mu_B = \sqrt{4 \pi
\alpha} / 2 m_e$ is the Bohr magneton, with $\alpha$ the electromagnetic fine
structure constant and $m_e$ the electron mass). The magnetic moment-induced
neutrino--nucleus scattering cross section is~\cite{Vogel:1989iv}
\begin{align}
  \frac{d\sigma_{\mu}(\nu N \to \nu N)}{dE_\text{rec}} = \mu_\nu^2 \alpha Z^2 F^2(E_\text{rec})
    \bigg(\frac{1}{E_\text{rec}} - \frac{1}{E_\nu}\bigg) \,,
  \label{eq:sigma-mm-N}
\end{align}
with the notation $E_\nu$ for the neutrino energy and $E_\text{rec}$ for the nuclear
recoil energy. Note that $d\sigma_\mu / dE_\text{rec}$ is proportional to the
square of the nuclear charge $Z$ because at low $E_\text{rec}$ the scattering is coherent
on all protons in the nucleus.  At higher $E_\text{rec}$, loss of coherence is taken into account by
the nuclear form factor $F(E_\text{rec})$~\cite{Harnik:2012ni}.  The neutrino--electron
scattering cross section is given by eq.~\eqref{eq:sigma-mm-N} with $Z$ and
$F(E_\text{rec})$ set to 1.

\begin{figure}
  \begin{center}
    \includegraphics[width=0.48\textwidth]{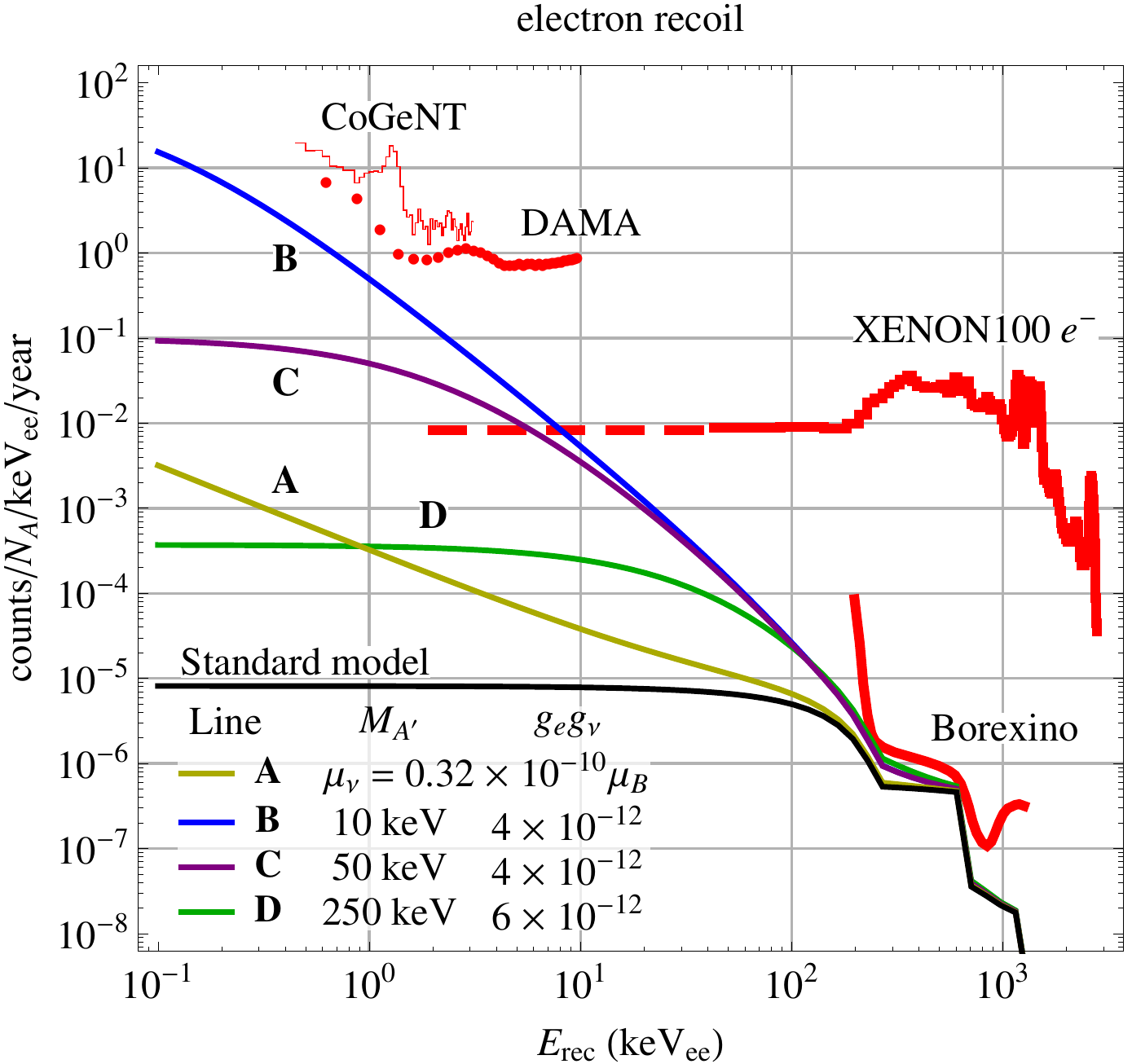}
    \hfill
    \includegraphics[width=0.48\textwidth]{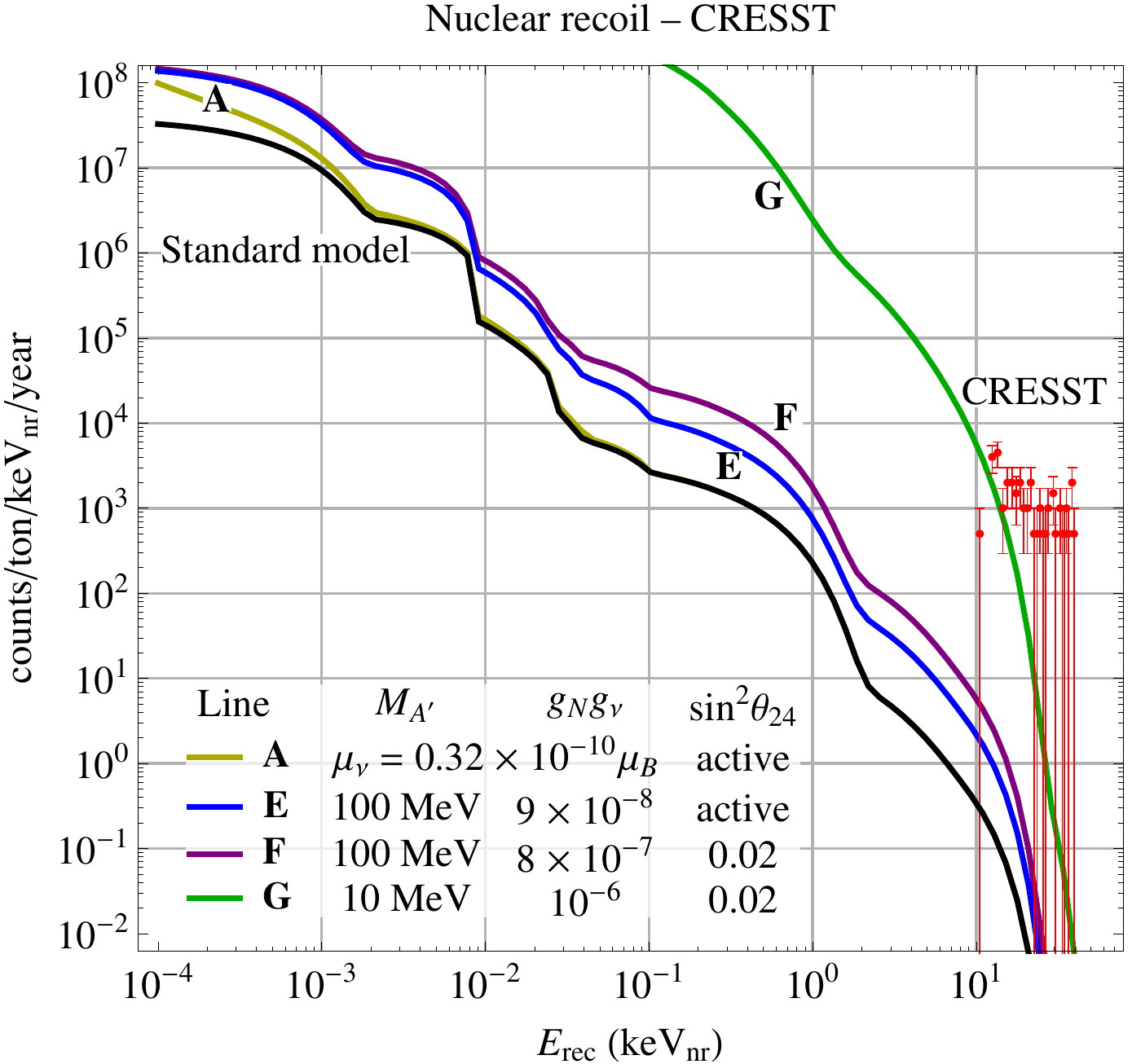}
  \end{center}
  \caption{Neutrino--electron scattering rates (left) and neutrino--nucleus
    scattering rates for germanium (right) in the Standard Model (black) and in
    various models with extended neutrino sectors (colored lines).  The curves
    labeled (A) illustrate the effect of a neutrino magnetic moment $\mu_\nu =
    0.32\times 10^{-10} \mu_B$; curves (B, C, D) are for a model where neutrino
    scattering is enhanced by exchange of a new $U(1)'$ gauge boson with mass
    $M_{A'}$ and couplings to neutrinos ($g_\nu$) and electrons ($g_e$) or
    nucleons ($g_N$) as given in the legend. This is, for instance, realized in
    scenarios with a sterile neutrino charged under $U(1)'$, and coupled to
    Standard model particles through kinetic mixing of the $U(1)'$ gauge boson
    and the photon. Curve (E) shows model with active neutrino--nucleus
    scattering through a light $A'$ boson (for instance a $U(1)_{B-L}$ model),
    and curves (F) and (G) illustrate a situation in which 2\% of solar neutrinos
    oscillate into a ``sterile'' state $\nu_s$, which is, however, charged under
    a $U(1)_B$ gauge group broken at a low scale~\cite{Pospelov:2011ha}.  We
    compare the model predictions to observed event rates in various experiments
    (red curves and data points)~\cite{Aprile:2011vb, Borexino:2011rx,
    Bernabei:2008yi, Aalseth:2011wp, Ahmed:2010wy}. Figure taken from
    ref.~\cite{Harnik:2012ni}.}
  \label{fig:dRdE}
\end{figure}

The differential neutrino--electron and neutrino--nucleus scattering rates
in the magnetic moment scenario are shown as a function of $E_\text{rec}$ in
fig.~\ref{fig:dRdE} (curves labeled ``(A)''). They exhibit the increase
with $1 / E_\text{rec}$ expected from eq.~\eqref{eq:sigma-mm-N} at low energy,
which leads to an enhancement of more than one order of magnitude at
1~keV recoil energy compared to the Standard Model.

\subsection{Sterile neutrinos and a light $U(1)'$ gauge boson}
\label{sec:sterile-model}
%-------------------------------------------------------------

An even larger enhancement of neutrino scattering rates in dark matter detectors
occurs in scenarios featuring, in addition to the massless photon, another
relatively light ($\ll 1$~GeV) gauge boson (``dark photon''). Since couplings of such gauge
bosons to Standard Model particles are strongly
constrained~\cite{Adelberger:2006dh, Jaeckel:2010ni, Harnik:2012ni}, the most
natural scenario of this type is one in which, in addition to the three active
neutrinos, a sterile neutrino sector exists which is uncharged under the
Standard Model gauge group, but couples to the new $U(1)'$ gauge force.  The
sterile sector can communicate with the Standard Model sector through a kinetic
mixing term of the form
\begin{align}
  \mathcal{L}_\text{mix} = -\frac{1}{2} \eps F'_{\mu\nu} F^{\mu\nu} \,,
  \label{eq:Lmix}
\end{align}
where $F_{\mu\nu}$ and $F'_{\mu\nu}$ are the electromagnetic and
$U(1)'$ field strength tensors, respectively, and $\eps$ is the small mixing
parameter. (An alternative possibility is a $U(1)'$ gauge boson which couples only to
the sterile neutrinos and Standard Model quarks, but not leptons~\cite{Pospelov:2011ha,
Pospelov:2012gm}.)

The sterile neutrinos can be produced in the Sun through their small mixing
with the active species, and they could scatter on nuclei in a terrestrial
detector with differential cross section
\begin{align}
  \frac{d\sigma_{A'}(\nu N \to \nu N)}{dE_\text{rec}} =
      \frac{4 \pi Z^2 \alpha \alpha' \eps^2 m_N F^2(E_\text{rec})}
           {E_\nu^2 (M_{A'}^2 + 2 E_\text{rec} m_N)^2}
      \big[ 2 E_\nu^2 + E_\text{rec}^2 - 2 E_\text{rec} E_\nu - E_\text{rec} m_N \big] \,,
  \label{eq:Aprime-xsec}
\end{align}
where $\alpha$ and $\alpha'$ are the electromagnetic and $U(1)'$ fine structure
constants, $m_N$ is the nuclear mass, $M_{A'}$ is the mass
of the new gauge boson, and the other quantities are defined as in
eq.~\eqref{eq:sigma-mm-N}.  The expression for neutrino--electron scattering is
obtained by the replacements $Z \to 1$, $F(E_\text{rec}) \to 1$, and $m_N \to m_e$
in eq.~\eqref{eq:Aprime-xsec}.

Possible signals of sterile neutrino scattering in dark matter detectors
through new $U(1)'$ gauge bosons are shown in fig.~\ref{fig:dRdE}, curves
(B)--(G).  (See plot legend for more details.) We see that these models can
easily lead to neutrino signals several orders of magnitude stronger than the
Standard Model prediction and thus highly relevant even to current dark matter
searches: experiments like Xenon-100 can be extremely sensitive to new physics
in the neutrino sector, which significantly enhances their physics case, but
can also constitute a problem because often neutrino signals cannot be easily
distinguished from genuine dark matter signatures.

This observation raises the question whether sterile neutrinos can explain some
of the anomalous signals reported from the DAMA~\cite{Bernabei:2008yi},
CoGeNT~\cite{Aalseth:2011wp}, and CRESST~\cite{Angloher:2011uu}.  As curve (B)
in the left panel of fig.~\ref{fig:dRdE} shows, a fraction of the excess events
observed in CoGeNT can indeed be accounted for if interpreted in terms of
electron recoils and if very conservative assumptions are made on the
sensitivity of Xenon-100 to low-energy electron recoils.
%Note that recently updated
%background estimates for CoGeNT suggest that, indeed, only a fraction of the
%observed low-energy excess events can have a non-standard origin~\cite{Aalseth:2012if}, and
%a recent measurement of the scintillation light yield from electron recoils in
%Xenon-100 indicates a drop in sensitivity at low energies~\cite{Aprile:2012an}.
%Still, determining conclusively whether a consistent explanation of CoGeNT
%an Xenon-100 is possible in models with an extended neutrino sector would
%require a dedicated study.
The CRESST signal could also be partially explained, but note that the energy
spectrum of CRESST events cannot be well reproduced, and that models like the
one shown in fig.~\ref{fig:dRdE} right, curve (D), are constrained by
the total event rate in DAMA (see fig.~4 in ref.~\cite{Harnik:2012ni}).
The annual modulation signal observed in DAMA will be discussed in the next
section.

%------------------------------------------------------------------------------
\section{Temporal modulation of neutrino signals}
%------------------------------------------------------------------------------

Dark matter scattering rates are expected to vary during the year because
the Earth's velocity with respect to the Milky Way's dark matter halo is
larger in summer than in winter~\cite{Freese:1987wuFig8}.  However, also neutrino
signals from the Sun are expected to modulate with time.

The simplest source of modulation is the varying Earth--Sun distance caused by
the ellipticity of the Earth's orbit. It leads to a roughly 3\% larger expected
count rate in winter than in summer, which corresponds to a 180$^\circ$ phase
shift compared to the commonly considered modulation signals from dark matter.
It is, however, interesting to note, that also dark matter scattering rates can
peak in winter in the higher energy parts of the recoil
spectrum~\cite{Freese:1987wuFig8}. Thus, in a detector with a high energy
threshold ($\gg 10$~keV), a dark matter modulation signal could be easily
mimicked by neutrinos from the Sun.

\begin{figure}
  \begin{center}
    \includegraphics[width=0.8\textwidth]{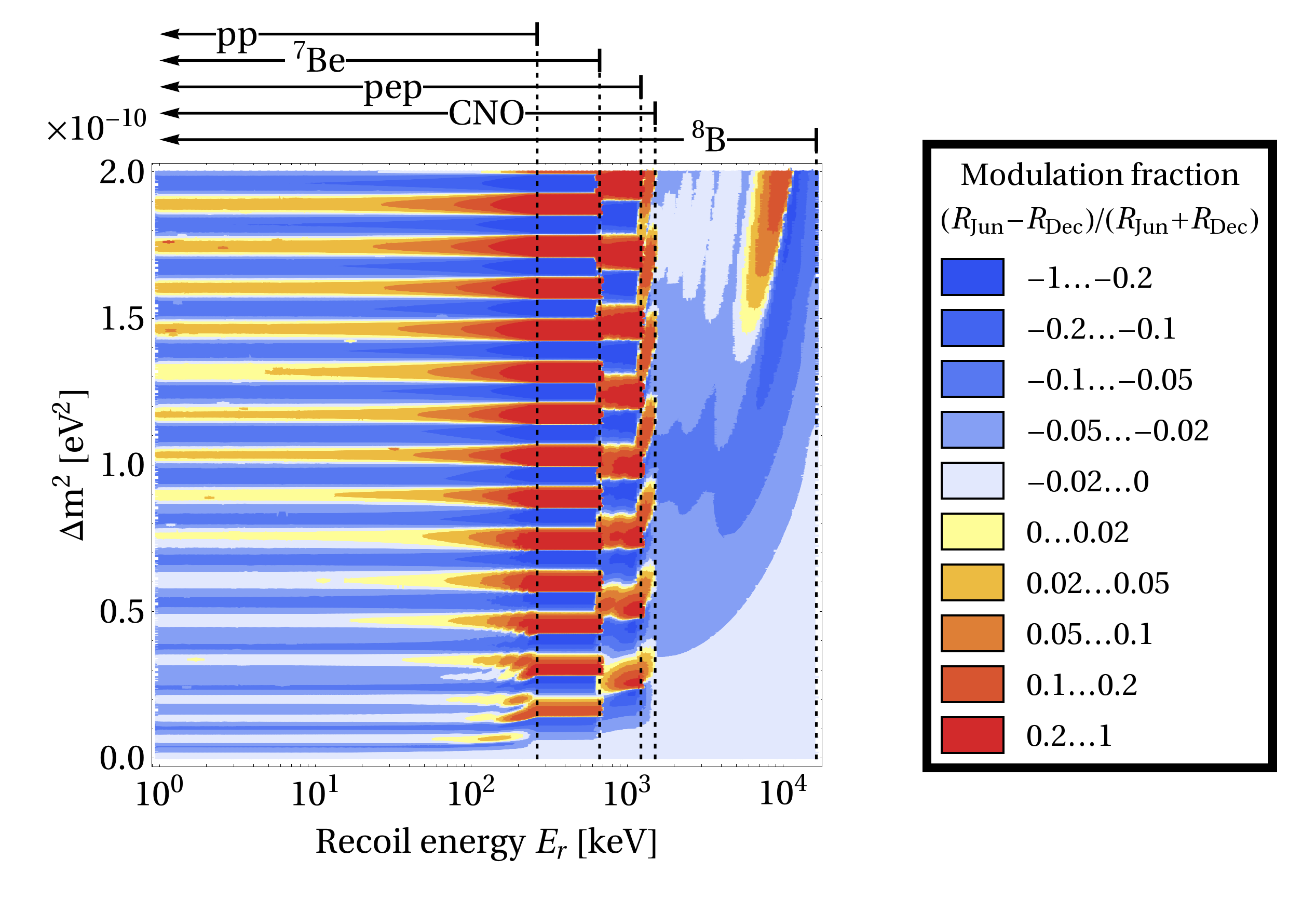}
  \end{center}
  \caption{Relative annual modulation $(R_{\rm Jun} - R_{\rm Dec})
    / (R_{\rm Jun} + R_{\rm Dec})$ for neutrino--electron scattering
    events in the $U(1)'$ model from sec.~\ref{sec:sterile-model}
    as a function of the recoil energy $E_\text{rec}$ and the mass squared difference
    between the mostly sterile mass eigenstate and $\nu_2$.
    Here $R_\text{Jun}$ and $R_\text{Dec}$ denote the differential
    count rate in events per keV at the summer and winter solstices,
    respectively. We have worked in a 2-flavor framework here.
    The vertical dashed lines indicate the cutoff energies of the
    different components of the solar neutrino spectrum. Figure taken from
    ref.~\cite{Harnik:2012ni}.}
  \label{fig:just-so}
\end{figure}

However, depending on the details of the model, neutrino signals in dark
matter detectors can also modulate with a different phase and amplitude.
This happens in particular in the model discussed in sec.~\ref{sec:sterile-model},
where the fake dark matter signal is due to sterile neutrinos
produced through oscillations (see also~\cite{Pospelov:2011ha}).
The sterile neutrino fraction in the solar neutrino flux can be significantly
larger in summer than in winter if the oscillation length is not too different
from the Earth--Sun distance, i.e.\ if the mass splitting $\Delta m^2$ between the mostly
sterile mass eigenstate and the mostly active mass eigenstate $\nu_2$ which
dominates the solar neutrino flux, is of order $10^{-10}$~eV$^2$. We illustrate
this in figure~\ref{fig:just-so} by plotting the fractional annual modulation
as a function of the recoil energy and of $\Delta m^2$. We see that large
modulation peaking in summer can occur in large parts of the parameter
space, i.e.\ no extreme fine-tuning is necessary to produce a dark matter-like
signal. Note that, even in this case, the modulation
phase would differ from the one expected from dark matter by about a month,
i.e.\ a high statistics measurement of annual modulation would still be
able to discriminate between dark matter and neutrino signals.

Other sources of temporal modulation of solar neutrino signals are
Mikheyev-Smirnov-Wolfenstein type matter effects~\cite{Mikheyev:1986wj,
Wolfenstein:1977ue} in the Earth (this typically requires more than one sterile
neutrino), sterile neutrino absorption in the Earth, and angle-dependent
detection efficiencies (see~\cite{Harnik:2012ni} for details on these
modulation mechanisms).

%------------------------------------------------------------------------------
\section{Constraints on dark photons and sterile neutrinos}
%------------------------------------------------------------------------------

\begin{figure}
  \begin{center}
    \includegraphics[width=\textwidth]{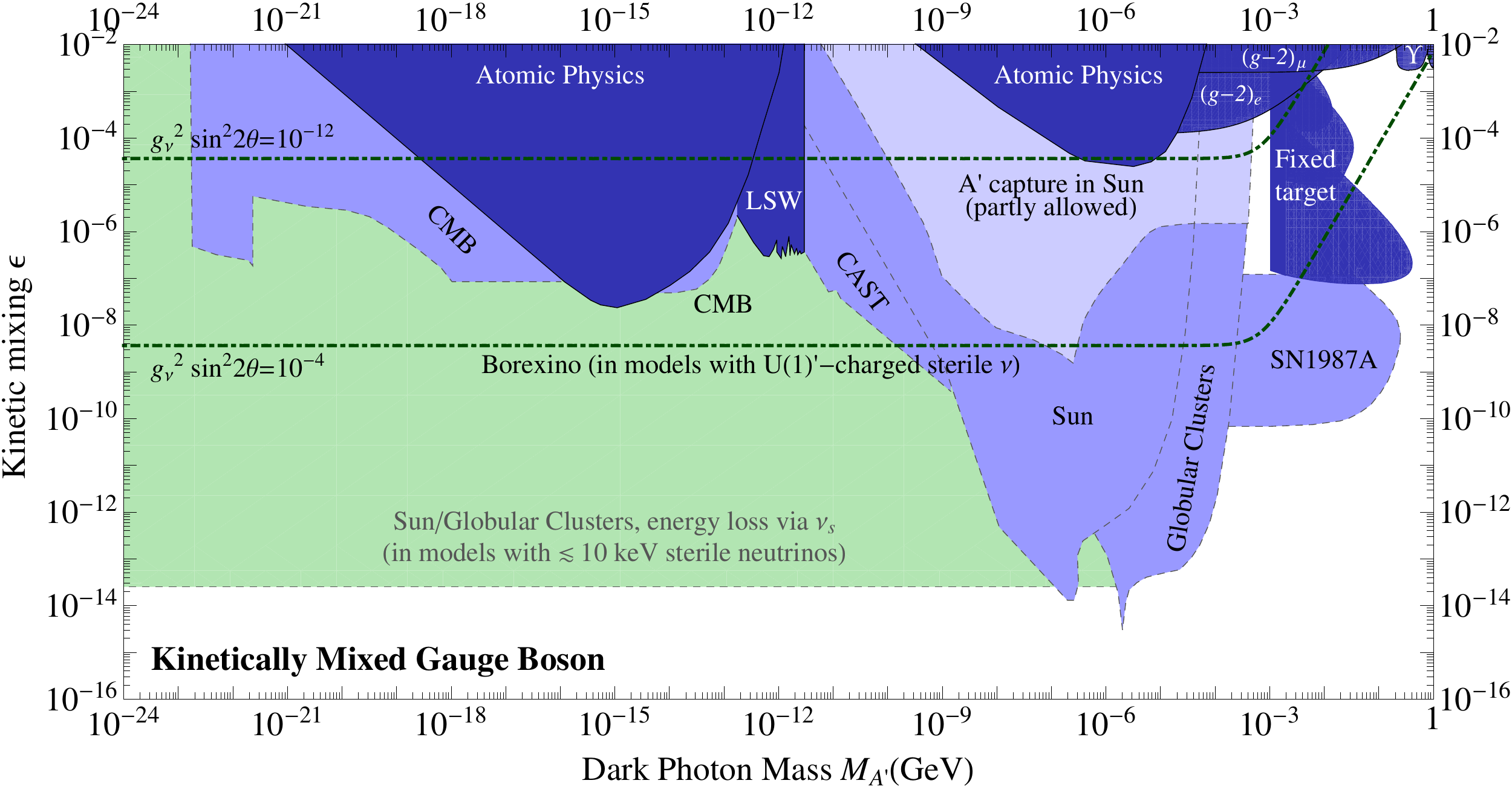}
  \end{center}
  \caption{Constraints on the mass $M_{A'}$ and the kinetic mixing parameter
    $\epsilon$ (see eq.~\eqref{eq:Lmix}) of a ``dark photon''. Most limits are
    taken from ref.~\cite{Jaeckel:2010ni}. See text, as well as
    ref.~\cite{Harnik:2012ni} and references therein for further details.  Figure
    taken from ref.~\cite{Harnik:2012ni}.}
  \label{fig:paramspace-Aprime}
\end{figure}

In sec.~\ref{sec:sterile-model}, we have only discussed specific benchmark
points for the models under consideration, but it is, of course, important
to investigate the full available parameter space. Indeed, a plethora of
constraints exist on light gauge bosons (dark photons) and on hidden sector
particles coupled to them. Here, we discuss constraints
for the specific case of a dark photon coupled to the Standard Model
through a kinetic mixing term of the form~\eqref{eq:Lmix}. These constraints
are summarized in fig.~\ref{fig:paramspace-Aprime}. (For constraints
on other types of $U(1)'$ gauge bosons, see ref.~\cite{Harnik:2012ni}
and references therein.)

At large dark photon mass $M_{A'}$, constraints come mainly from particle
physics experiments, in particular fixed target (beam dump) experiments,
measurements of the anomalous magnetic moment $g-2$ of the electron and
the muon, and $\Upsilon$ decays studied at $B$ factories~\cite{Bjorken:2009mm,
Batell:2009di, Essig:2009nc, Essig:2010gu}. (Note that a dark
photon with parameters just below the region labeled $(g-2)_\mu$ could
explain the observed discrepancy between the measured muon anomalous
magnetic moment and the Standard Model prediction~\cite{Bennett:2006fi}.)

For $M_{A'}$ in the eV--MeV region, very strong limits can be derived from
stellar evolution: dark photons would provide a very efficient energy loss
mechanism for stars and supernovae~\cite{Jaeckel:2010ni, Redondo:2008aa,
Dent:2012mx}, significantly shortening their lifetime (for stars) or reducing
their energy output (for supernovae).  The resulting limits are especially
strong if $M_{A'}$ is similar to the typical thermal energies in the
star~\cite{Redondo:2008aa}, so that $\eps$ is constrained down to $10^{-14}$ in
this regime. Note, however, that dark photons with stronger couplings
(indicated by the light blue region in fig.~\ref{fig:paramspace-Aprime}) cannot
be ruled out this way because they would not be able to leave the
star/supernova due to absorption~\cite{Redondo:2008aa}.  Limits could still be
obtained from stellar evolution constraints on anomalous heat transport inside
the star~\cite{Raffelt:1988rx}, but for $M_{A'} \gtrsim 100$~keV, we expect
these constraints to disappear as well~\cite{Raffelt:1988rx, Harnik:2012ni}.

For sub-eV dark photon masses, the strongest constraints are obtained from
helioscopes like CAST~\cite{Redondo:2008aa}, from ``Light Shining through
Walls'' (LSW) experiments~\cite{Ahlers:2007qf}, from tests of the Coulomb law
in atomic physics~\cite{Bartlett:1988yy}, and from distortions of the CMB
spectrum~\cite{Mirizzi:2009iz}.

In models which contain in addition to the dark photon also one or several
light sterile neutrinos (like the scenario discussed in
sec.~\ref{sec:sterile-model}), additional constraints arise from the fact that
the sterile neutrino(s) would act as ``minicharged
particles''~\cite{Davidson:2000hf, Jaeckel:2010ni}. The strongest limits in
this case arise again from stellar cooling (light green region in
fig.~\ref{fig:paramspace-Aprime}) and from Borexino~\cite{Harnik:2012ni,
Borexino:2011rx}.

%------------------------------------------------------------------------------
\section{Summary}
%------------------------------------------------------------------------------

To summarize, we have shown that rather minimal extensions of the neutrino
sector of the Standard Model can lead to very interesting signals in dark
matter direct detection experiments. As examples, we have considered scenarios
with neutrino magnetic moments and models with a sterile neutrino sector
enjoying a new gauge symmetry broken at a scale $\ll 1$~GeV. We have
shown that very large neutrino--electron and neutrino--nucleus scattering
rates, well within the reach of current dark matter detectors, can occur at
$\mathcal{O}(10\text{ keV})$ recoil energy.  At the higher energies probed by
dedicated neutrino detectors, no anomalous signals would be expected.

On the one hand, neutrino signals in dark matter detectors enhance the physics
case of these experiments by allowing them to probe additional types of new
physics besides dark matter. However, they can also constitute a problem
because, especially for small event samples, they can be easily confused with
genuine dark matter signals.  For instance, we have shown that neutrino signals
can exhibit temporal modulation similar to the one expected from dark matter.

%------------------------------------------------------------------------------
\section*{References}
%------------------------------------------------------------------------------
%\begin{center}
%  \rule{10cm}{0.25pt}
%\end{center} 
%\vspace{-0.7cm}
\footnotesize
\bibliographystyle{iopart-num}
\bibliography{new-signals}

\end{document}